\documentclass[twocolumn,letter]{jpsj3}
\usepackage{color}
\usepackage{braket}
\usepackage{mathrsfs}

\def\runauthor{Y. Akagi and Y. Motome}

\title{Spin Chirality Ordering and Anomalous Hall Effect \\in the Ferromagnetic
Kondo Lattice Model on a Triangular Lattice}

\author{Yutaka Akagi\thanks{E-mail address: akagi@aion.t.u-tokyo.ac.jp} and Yukitoshi Motome 
}
\inst{Department of Applied Physics, University of Tokyo, 
Hongo 7-3-1, Bunkyo-ku, Tokyo 113-8656
}
\abst{
We investigate the ground-state phase diagram for 
the ferromagnetic Kondo lattice model on a triangular lattice 
by a variational calculation for various spin orderings up to a four-site unit cell.  
We find that a noncoplanar four-sublattice ordering with a finite scalar spin chirality emerges 
at and around 1/4 filling, 
in addition to the 3/4-filled case, which was predicted 
to be induced by the perfect nesting of the Fermi surface
[I. Martin and C. D. Batista: Phys. Rev. Lett. {\bf 101} (2008) 156402]. 
The 1/4-filling phase is stable in a wider range of parameters 
than the 3/4-filling one, 
and includes a large region of gapped insulating state 
characterized by a Chern number. 
We also compute the Hall conductivity in the chiral-ordered phases.
}

\kword{spin chirality, anomalous Hall effect, Kondo lattice model, frustration, triangular lattice, Berry phase, Chern number}

\begin{document}
\maketitle

Spin chirality has been an issue of interest 
as an emergent composite degree of freedom  
in a broad range of condensed matter physics.\cite{Villain1977,Miyashita1984,Wen1989,Kawamura1992}
Noncollinear spin configurations give rise to a finite spin chirality, e.g., 
the vector spin chirality $\vec{S}_1 \times \vec{S}_2 + \vec{S}_2 \times \vec{S}_3 + \vec{S}_3 \times \vec{S}_1$ 
or the scalar spin chirality $\vec{S}_1 \cdot (\vec{S}_2 \times \vec{S}_3)$, 
defined on a triangle of spins. 
The latter scalar chirality, which becomes finite for a noncoplanar spin configuration,  
has recently drawn considerable attention 
not only in localized spin systems but also in itinerant electron systems 
as an intrinsic origin of the anomalous Hall effect (AHE). 
When itinerant electrons are coupled to such specific spin configuration, 
electrons experience an internal magnetic field according to the solid angle spanning three spins 
through the so-called Berry phase,  
which can result in AHE. 
This new mechanism was first theoretically proposed for a chiral order 
in the ferromagnetic Kondo lattice model on a two-dimensional kagome lattice,\cite{Ohgushi2000} 
and developed in a three-dimensional frustrated system.\cite{Shindou2001}
The relevance has been discussed in experiments on metallic pyrochlore oxides.\cite{Taguchi2001}

Recently, this chirality mechanism was explored for 
triangular lattice systems. 
Martin and Batista pointed out the possibility of AHE in  
a Kondo lattice model on the triangular lattice 
at 3/4 filling of itinerant electrons, 
which is brought about by a noncoplanar four-sublattice spin ordering 
due to the perfect nesting of the Fermi surface.\cite{Martin2008} 
This proposal is attractive since similar noncoplanar four-sublattice orders are encountered 
in some triangular-lattice spin systems.\cite{Mn_mono,2D_He3} 

Although previous studies have revealed a nontrivial relation 
between chirality ordering and AHE, 
they rely on the assumption that the spin configuration is a given texture and 
not affected by the coupling to itinerant electrons. 
In spin-charge coupled systems, however, 
the spin state is determined in a self-consistent manner 
with the itinerant electron state through the interplay between spin and charge.
Hence, it remains unclear whether such chiral ordering is realized under spin-charge interplay. 
For further understanding of spin-chirality-related phenomena, 
it is necessary to carry out a microscopic analysis  
of the most stable spin state 
and to clarify whether the chiral order with AHE remains stable 
even when considering the interplay between spin and charge.

In this study, we investigate the ground state of the ferromagnetic Kondo lattice model 
on the frustrated triangular lattice 
with emphasis on the possibility of chiral ordering. 
We compare the energies for various spin configurations up to four-sublattice orders, 
and determine the ground-state phase diagram. 
As a result, we find that a noncoplanar four-sublattice spin ordering with a finite scalar spin-chirality 
emerges around 1/4 filling, in addition to the 3/4 filling predicted in the previous study.\cite{Martin2008} 
This new phase includes both metallic and insulating phases and is stabilized in a wider parameter region 
than the 3/4-filling phase.
We also discuss AHE in these chiral-ordered phases by calculating the Hall conductivity, which is quantized in the insulating phases corresponding to the associated Chern number.\cite{Shindou2001,Martin2008}

We consider the ferromagnetic Kondo lattice model on the triangular lattice
as one of the fundamental models for describing the interaction between localized spins and conduction electrons under frustration.
The Hamiltonian is given by
\begin{align} 
{\cal H}=&-t\sum_{\langle i,j \rangle,\alpha } ( c^{\dagger}_{i,\alpha } c_{j,\alpha }+\mathrm{h.c.}) \notag\\
&-J_H \sum_{i,\alpha ,\beta }
    c^{\dagger}_{i,\alpha } \vec {\sigma}_{\alpha \beta }  c_{i,\beta } \cdot \vec {S_i} 
   +J_K \sum_{\langle i,j\rangle} \vec {S_i} \cdot \vec {S_j},
\end{align} 
where $c^{\dagger}_{i,\alpha }$($c_{i,\alpha }$) is a creation (annihilation) operator for a conduction electron with spin $\alpha$ on site \textit{i}, 
and \textit{t} is the transfer integral, $J_H$ is the Hund's-rule coupling, $\vec {\sigma}_{\alpha \beta }=({\sigma}^x_{\alpha \beta },{\sigma}^y_{\alpha \beta },{\sigma}^z_{\alpha \beta })$
 is a vector of Pauli matrices, $\vec {S_i}$ is a localized spin on site \textit{i}, and $J_K$ is the antiferromagnetic superexchange interaction between localized spins. 
 We consider classical spins for $\vec {S_i}$ with $|\vec {S_i}|=1$.
 The sums $\langle i,j\rangle$ are taken over the nearest-neighbor sites on the triangular lattice. Hereafter, we take $t=1$ as an energy unit.
 
We investigate the ground state of the model given by eq. (1) while varying the electron density $n=\frac{1}{N}\sum_{i\alpha }\langle c^{\dagger}_{i,\alpha } c_{i,\alpha }\rangle$ (\textit{N} is the total number of sites), $J_H$, and $J_K$. 
\begin{figure}[t]
\begin{center}
\includegraphics[width=8.0cm]{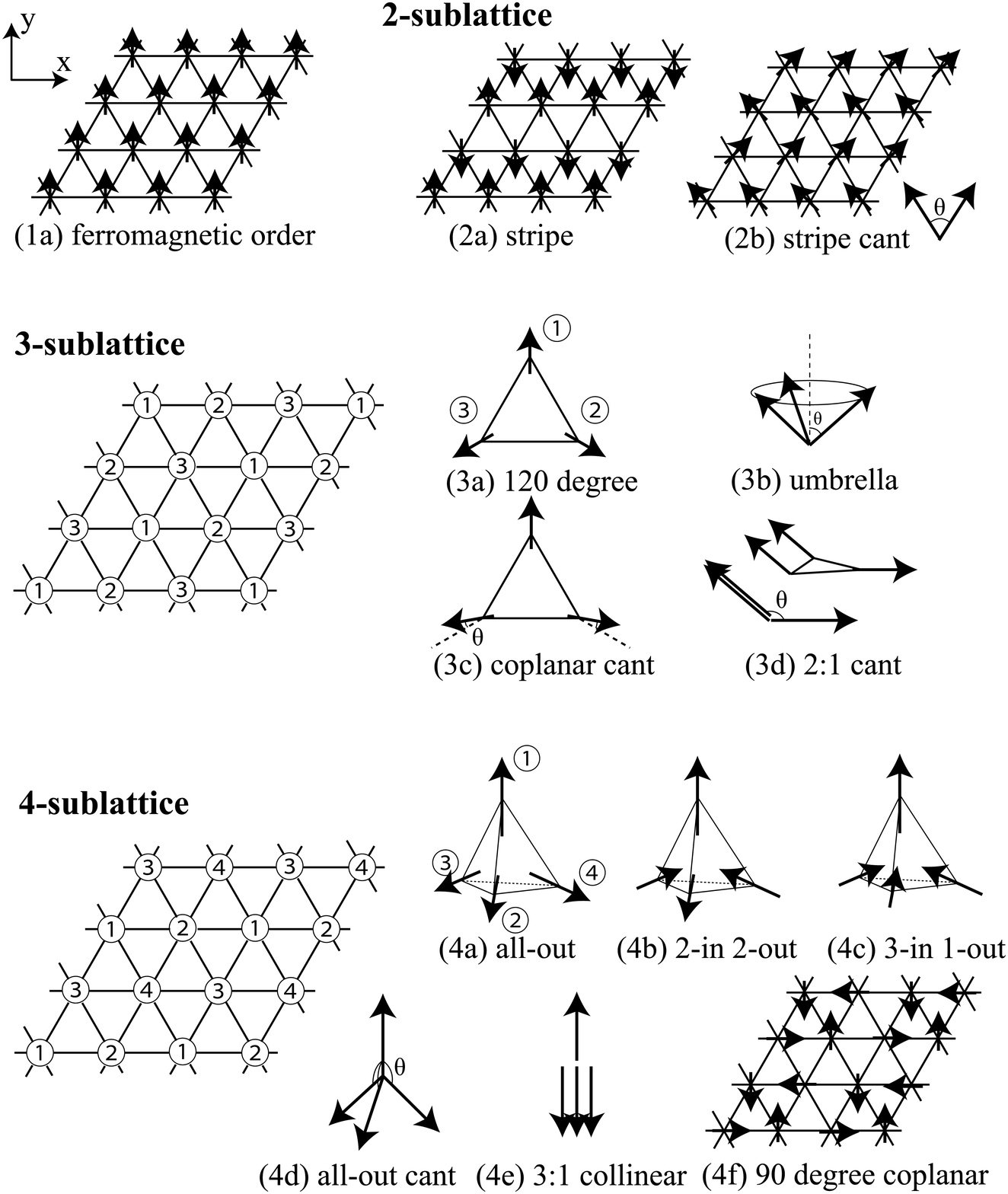}
\end{center}
\caption{Ordering patterns of localized spins used in the variational calculations:
(1a) ferromagnetic order, (2a) and (2b) two-sublattice orders, (3a)-(3d) three-sublattice orders, and (4a)-(4f) four-sublattice orders. 
See text for details.
}
\label{all_order}
\end{figure}
We compare the ground-state energies for different ordered states of the localized spins and determine the most stable ordering by a variational calculation.
 In the calculation, we consider 13 different types of ordered states, up to a four-sublattice unit cell, as shown in Fig. \ref{all_order}.
Figure \ref{all_order}(1a) shows a ferromagnetic order.
 Figures \ref{all_order}(2a) and 
\ref{all_order}(2b) show two-sublattice orders: (2a) a collinear stripe order and (2b) a stripe order with a canting angle $\theta $. 
 Figures \ref{all_order}(3a)-\ref{all_order}(3d) show three-sublattice orders: (3a) a $120^{\circ }$ noncollinear order,
 (3b) a noncoplanar umbrella-type order with angle $\theta $ 
 (canted in the normal direction to the coplanar plane from the $120^{\circ }$ order),
 (3c) a coplanar order with canting angle $\theta $ for two spins from $120^{\circ }$ order,
 and (3d) a 2:1-type order with two parallel spins that have angle $\theta $ to the remaining one.
 Figures \ref{all_order}(4a)-\ref{all_order}(4f) show four-sublattice orders:
 (4a) an all-out-type order which was discussed in Ref. 8, 
(4b) a two-in two-out-type order,
 (4c) a three-in one-out-type order,
 (4d) an all-out-type order with canting angle $\theta $ for three spins,
 (4e) a 3:1 collinear order, 
 and (4f) a coplanar order with a $90^{\circ }$ flux-type configuration.
For the given parameters \textit{n}, $J_H $, and $J_K $, we determine which spin state is most energetically stable by comparing the ground-state energies 
for all the spin states listed above.
For the states (2b), (3b), (3c), (3d), and (4d), we optimize the canting angle $\theta $. 
Note that 
(2b) with $\theta =\pi $, (3b) with $\theta =\frac{\pi }{2}$, (3c) with $\theta =0$, 
(4d) with $\theta = \cos^{-1}(-1/3)$, $\theta = \cos^{-1}(+1/3)$, and $\theta = \pi$ are equivalent to (2a), (3a), (3a), (4a), (4c), and (4e), respectively.
Later, to check the local stability of (4a), 
we will also try a general configuration with all possible spin directions 
within the four-site unit cell.
Although an incommensurate order might take place for a general filling, 
we focus on uniform $\vec{q}=0$ orders with the magnetic unit cells listed above, 
particularly, on their stability at and around commensurate fillings, as we discuss below.

 \begin{figure}[t]
 \begin{center}

  \includegraphics[width=80mm]{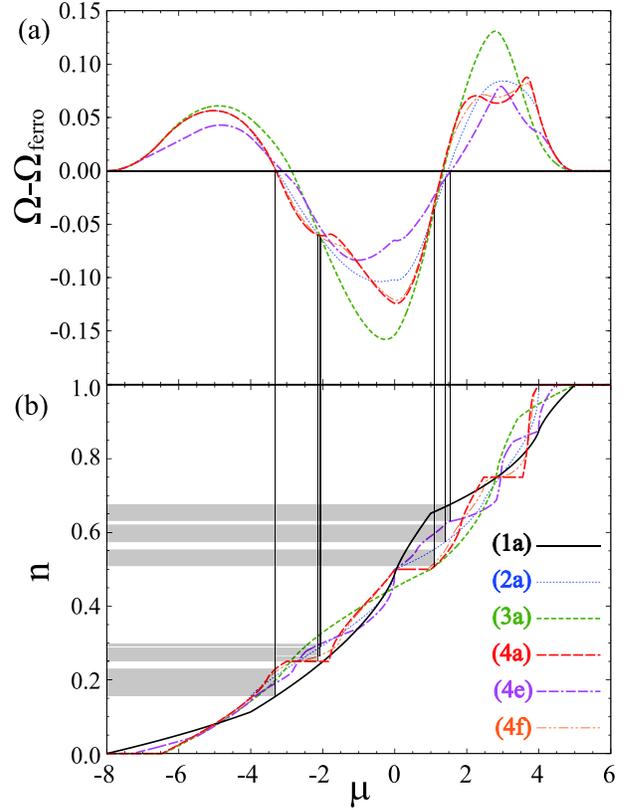}
 \end{center}
 \caption{(Color online) (a) Grand-canonical potential at $T=0$, $\Omega = \langle {\cal H} \rangle - \mu n$,  
for relevant ordered states measured from that for the ferromagnetic order and (b) the electron density 
 $n$ as a function of the chemical potential $\mu $. The data are for $J_H=2.0$ and $J_K=0$.
Phase-separated regions, shaded areas in (b), 
are determined from the jumps of $n$ in (b) 
associated with the crossing points in (a). 
}
 \label{phase-separation}
\end{figure}
Figure \ref{phase-separation}(a)
 shows a comparision of the grand-canonical potential at $T=0$, $\Omega = \langle {\cal H} \rangle - \mu n$, for various ordered states as a function of the chemical potential $\mu $  
 for $J_H=2.0$ and $J_K=0$, as an example. 
The data are computed by approximating the integral over the (unfolded) first Brillouin zone using the sum over grid points of $1600 \times 1600$. 
We here plot $\Omega$ for relevant ordered states at these parameters measured from that for the ferromagnetic state in Fig. \ref{all_order}(1a). 
(All other states not shown have higher $\Omega$.) 
In the low- and high-$\mu $
 regions, the ferromagnetic state (1a) gives the lowest $\Omega$. 
However, in the intermediate region, $-3.3\hspace{0.3em}\raisebox{0.4ex}{$<$}\hspace{-0.75em}\raisebox{-.7ex}{$\sim $}\hspace{0.3em}\mu \hspace{0.3em}\raisebox{0.4ex}{$<$}\hspace{-0.75em}\raisebox{-.7ex}{$\sim $}\hspace{0.3em}1.6$, other states have a lower $\Omega$. 
For each $\mu $
 range, the state that has the lowest $\Omega$ gives the ground state.
 
To find the ground-state phase diagram as a function of $n$, it is necessary to examine the relationship between $n$ and $\mu$. 
In general, a phase transition between different magnetic orders is of first order and accompanies a jump of \textit{n}.
This is demonstrated in Fig. \ref{phase-separation}(b), which plots \textit{n} as a function of $\mu $. 
For example, at $\mu \simeq -3.3$, the ground state changes from a ferromagnetic state [Fig. \ref{all_order}(1a)] to a four-sublattice all-out-type ordered state [Fig. \ref{all_order}(4a)], as shown in Fig. \ref{phase-separation}(a). At the same $\mu$, as shown in Fig. \ref{phase-separation}(b), the density $n$ has different values for these two states, i.e., $n\simeq 0.16$ for the former and $n \simeq 0.23$ for the latter, which results in a discontinuous change of $n$ at the phase transition. 
Since the system is unstable and cannot have a fixed density in the discontinuous regime, the jump of $n$ signals a phase separation. 
Similar analysis identifies phase-separated regions, as demonstrated in Fig. \ref{phase-separation}.

We perform the comparison of $\Omega$ 
and the identification of phase-separated regions while varying the parameters \textit{n}, $J_H$, and $J_K$. 
For $J_H \ll t$, in particular, at $J_K=0$, the difference in $\Omega$ among different orders becomes very small. In such cases, we determine the ground states using the perturbation theory 
 in $J_H/t$ up to fourth order.

    \begin{figure}[t]
 \begin{center}
  \includegraphics[width=8.5cm]{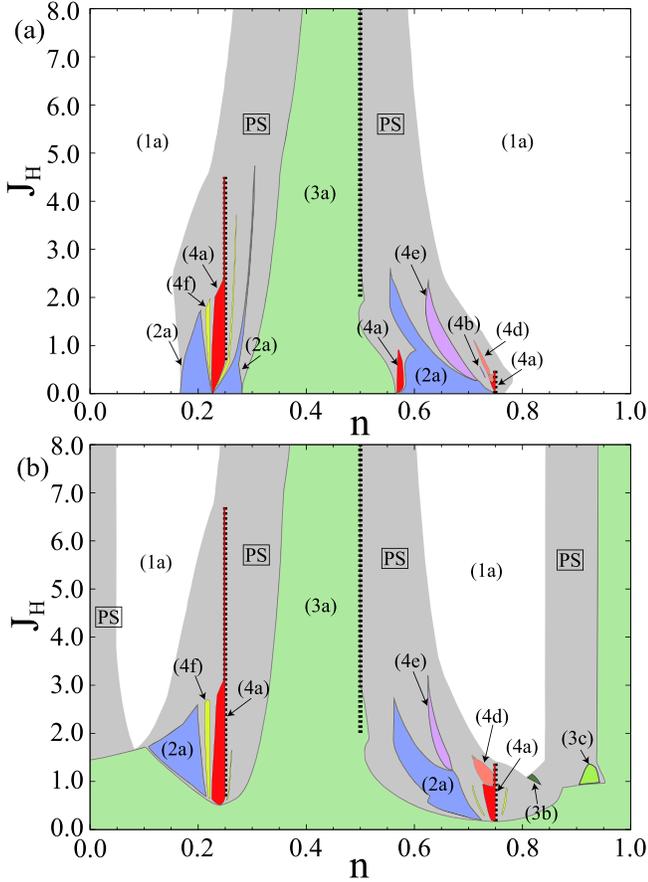}
 \end{center}
 \caption{(Color online) 
 Ground-state phase diagram for (a) $J_K=0$ and (b) $J_K=0.01$. 
The regions (1a)-(4f) correspond to the ordering patterns in Figs. \ref{all_order}(1a)-\ref{all_order}(4f), respectively.
 The vertical dashed lines at $n=1/4$, $1/2$, and $3/4$ show gapful insulating regions. PS indicates a phase-separated region. 
}
 \label{phase-diagram}
\end{figure}
   Figure \ref{phase-diagram}(a) shows the result of the phase diagram as functions of \textit{n} and $J_H$ at $J_K=0$.
   In the low- and high-density regions, a ferromagnetic metallic phase appears [Fig. \ref{all_order}(1a)]. The phase becomes wider as $J_H$ increases.
   This ferromagnetic phase is stabilized by the double-exchange mechanism\cite{Zener1951}.
On the other hand, a three-sublattice $120^{\circ }$ order emerges near half-filling $n\simeq 1/2$ 
[Fig. \ref{all_order}(3a)]. 
In particular, the system becomes insulating at $n=1/2$ for $J_H>2$.
 This is understood by considering the fact that, at half-filling, 
the second-order perturbation in \textit{t}$/J_H$ leads to an effective
 antiferromagnetic interaction between localized spins. 
Under the frustration in the triangular lattice, the antiferromagnetic interaction stabilizes the $120^{\circ }$ order.
In the large-$J_H$ region, a phase separation takes place between the ferromagnetic state and the $120^{\circ }$-ordered state.

For smaller $J_H$, the phase diagram becomes more complicated. 
Among various phases, the most interesting point is that a four-sublattice all-out order with a finite scalar chirality [Fig. \ref{all_order}(4a)]
 is realized near 1/4 filling and 3/4 filling.\cite{note} 
The phase at 3/4 filling is the one predicted by Martin and Batista 
as a consequence of the perfect nesting of the Fermi surface\cite{Martin2008}. 
Our energy comparison confirms the prediction. 
In contrast, the phase around 1/4 filling is not of nesting origin, since the Fermi surface is almost circular at this filling for $J_H=0$.
This is a new chiral phase with a four-sublattice order, which was not predicted in the previous work.
Because of the noncoplanar spin order, this chiral phase exhibits AHE as in the 3/4 filling phase; we will return to this point later.

As shown in Fig. \ref{phase-diagram}(a), the new phase at $n \simeq 1/4$ is stabilized in a wider range of parameters than the nesting-driven 3/4-filling phase; 
the former is stabilized for $0 < J_H \leq 4.5$, whereas the latter is limited for $0 < J_H \leq 0.51$. 
Moreover, the 1/4-filling phase becomes insulating 
in a wide $J_H$ range of $0.69\leq J_H\leq 4.5$. 
The gap opens between the lowest and second bands of the four bands 
under the four-sublattice order (each band has a twofold degeneracy in terms of spin).
This gap appears to robustly stabilize the chiral order; 
in fact, 
we confirm that  
the gapped chiral state remains 
stable even when allowing all possible spin configurations while changing the relative angle of spins continuously within the four-site unit cell.

When we switch on the superexchange interaction $J_K$, the chiral phases as well as the three-sublattice $120^{\circ }$-ordered phase become more stable.
 Energy changes are easily calculated from the last term in eq. (1), for example, 
 $-J_K$ ($-3J_K/2$) per site for the four-sublattice all-out 
 (three-sublattice $120^{\circ }$) ordered state.
 Figure \ref{phase-diagram}(b) shows the phase diagram at $J_K=0.01$. 
 Indeed, the $120^{\circ }$-ordered state covers a wider range of parameters; in particular, it extends to the small-$J_H$ region and $n\sim 1$ region. 
 In addition, the four-sublattice chiral phases around 1/4 and 3/4 fillings are also enlarged. 
 Instead, the ferromagnetic regions are reduced and surrounded by phase separations up to the $120^{\circ }$-ordered state. 
 Thus, $J_K$ stabilizes the chiral phases as well as the $120^{\circ }$-ordered state.

 \begin{figure}[t]
 \begin{center}

  \includegraphics[width=70mm]{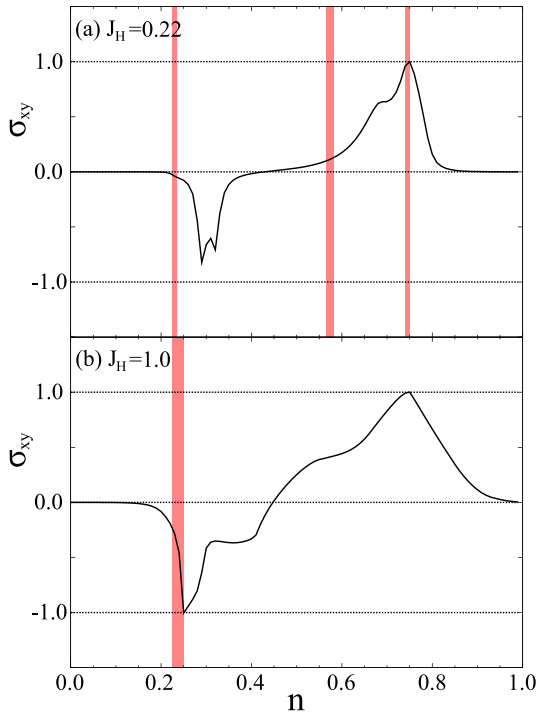}
  
 \end{center}
 \caption{(Color online) Hall conductivity calculated at $T=0$ as a function of the electronic density \textit{n} while assuming the four-sublattice all-out order [Fig. \ref{all_order}(4a)]
for (a) $J_H=0.22$ and (b) $J_H=1.0$. 
The shaded areas represent the regions where the four-sublattice all-out order is realized 
at $J_K=0$ 
in Fig. \ref{phase-diagram}(a).
}
 \label{sigma_xy}
\end{figure}

Let us discuss AHE for the model eq. (1), while focusing on the chiral phases near 1/4 and 3/4 fillings.
We calculate the Hall conductivity 
by the Kubo formula
\begin{align}
\sigma _{xy}&=
\sum_{n,\textbf{k}} \frac{f_\textit{F}(\epsilon _{n\textbf{k}})}{2\pi i} \!\!\! \sum_{m\neq n,\alpha \beta }\!\!\! \frac{J_{x; \textbf{k}}^{n_{\alpha }m_{\beta }} J_{y;  \textbf{k}}^{m_{\beta }n_{\alpha }}-J_{x; \textbf{k}}^{m_{\alpha }n_{\beta }} J_{y; \textbf{k}}^{n_{\beta }m_{\alpha }}}{(\epsilon _{n\textbf{k}}-\epsilon _{m\textbf{k}})^2} \notag\\
&=
\sum_n \int_{\mathrm{BZ}}\mathrm{tr}\; d\mathscr{A}_{n\textbf{k}} \frac{f_\textit{F}(\epsilon _{n\textbf{k}})}{2\pi i},
\end{align}
where 
$f_\textit{F}$ is the Fermi distribution function, 
$J_{\nu; \textbf{k}}^{n_{\alpha }m_{\beta }}=\bra{n_\alpha\textbf{k} }J_\nu\ket{m_\beta\textbf{k} }$ 
($J_\nu$ is a current operator in the direction $\nu=x, y$), 
and 
${\cal H}\ket{n_\alpha\textbf{k} } = \epsilon _{n\textbf{k}}\ket{n_\alpha\textbf{k} }$ with a degeneracy index $\alpha $. 
Here, we take $e^2/h=1$ ($e$ is the elementary charge and $h$ is the Planck constant).
The second line is obtained by considering the non-Abelian connection $\mathscr{A}_{n\textbf{k}}=\psi _{n\textbf{k}}^{\dagger}d\psi _{n\textbf{k}}$, which is given by an $M\times M$ matrix-valued one-form associated with a multiplet $\psi _{n\textbf{k}}=(\ket{n_1\textbf{k}},\cdot \cdot \cdot, \ket{n_M\textbf{k}})$.\cite{nonAbel1,nonAbel2,Chern_number}
The integral is taken over the first Brillouin zone. 
For the four-sublattice all-out order, 
the multiplet with $M=2$ describes the local \textit{SU}(2) symmetry in the \textbf{k}-space owing to the band degeneracy associated with spins. 
When the Fermi level is in the gap, it is known that the Hall conductivity is quantized and given by the summation of Chern numbers for the occupied bands; 
$
\sigma _{xy}= 
\sum_{n(\epsilon _{n\textbf{k}}\leq \epsilon _{\textit{F}})}C_n,
$
where the Chern number is defined as $C_n=\int_{\mathrm{BZ}}\mathrm{tr}\; d\mathscr{A}_{n\textbf{k}}f_\textit{F}(\epsilon _{n\textbf{k}})/(2\pi i)$\cite{Kohmoto1985}.

Figure \ref{sigma_xy} shows the result of $\sigma _{xy}$ obtained using eq. (2) while assuming the four-sublattice all-out order 
in the entire region of \textit{n} for (a) $J_H=0.22$ and (b) $J_H=1.0$. 
The shaded areas correspond to the regions where the four-sublattice all-out order is realized in the phase diagram at $J_K=0$ in Fig. \ref{phase-separation}(a).
The results show that $\sigma_{xy}$ has a finite value in these chiral states. 
The Chern numbers associated with four bands are given by $-1$, $1$, $1$, and $-1$ from the lowest band to the highest band, respectively.
Thus, the Hall conductivity is quantized as $\sigma _{xy}=-e^2/h$ at 1/4 filling and $\sigma _{xy}=+e^2/h$ at 3/4 filling
 when the system is a gapped insulator.\cite{Shindou2001,Martin2008}
This is observed in Fig. \ref{sigma_xy} at 3/4 filling for $J_H=0.22$ [Fig. \ref{sigma_xy}(a)]
 and at 1/4 filling for $J_H=1.0 $ [Fig. \ref{sigma_xy}(b)].

To summarize, we have clarified that a four-sublattice noncoplanar magnetic order 
is stabilized around 1/4 filling in the ferromagnetic Kondo lattice model on a triangular lattice. 
This phase is different from the 3/4-filling phase, which was predicted previously 
to appear owing to the perfect nesting of the Fermi surface. 
The region is wider than the 3/4-filling one, and includes both metallic and insulating states. 
It is further stabilized by the superexchange coupling between localized spins. 
The noncoplanar order retains a finite scalar spin chirality, resulting in the anomalous Hall effect; 
in particular, the Hall conductivity is quantized according to the Chern number in the insulating regions. 
Since our 1/4-filling phase is not directly related to nesting instability, 
it is expected to survive in a broad range of triangular-lattice systems, compared with the 3/4-filling phase. 
The stability against a modification of band structure is a problem for future study. 
It is also of interest to examine the stability against quantum and thermal fluctuations by, e.g., spin-wave analysis and Monte Carlo simulation. 
Our results will be relevant to spin-charge coupled phenomena in 
itinerant triangular-lattice compounds, such as delafossites.

\acknowledgements
{
We acknowledge helpful discussions with Takahiro Misawa, Masafumi Udagawa, and Youhei Yamaji. 
Y.A. gratefully thanks Atsuo Shitade for his fruitful comments.
This work was supported by Grants-in-Aid for Scientific Research 
(No. 19052008 and 21340090), 
Global COE Program ``the Physical Sciences Frontier", the Next Generation Super Computing 
Project, and Nanoscience Program, from MEXT, Japan.
}


\begin{thebibliography}{20}
\bibitem{note}
The four-sublattice order also appears at around $n=0.57$, whereas we do not focus on this state 
since it is suppressed by a small $J_K$, as seen in Fig. \ref{phase-diagram}(b).
\end{thebibliography}
\end{document}